\documentclass[aps,twocolumn]{revtex4}
\usepackage{mathrsfs}
\usepackage{amsmath}
\usepackage{graphicx}
\usepackage{subfigure}
\usepackage{epstopdf}

\begin{document}

\renewcommand{\bottomfraction}{.7}
\setcounter{topnumber}{12} \setcounter{bottomnumber}{12}
\setcounter{totalnumber}{20} \setcounter{dbltopnumber}{12}





\title{The chiral condensate in a constant electromagnetic field at ${\cal O}(p^6)$}

\author{Elizabeth S. Werbos}
\email{ewerbos@physics.umd.edu}

\affiliation{Department of Physics, University of Maryland,
College Park, MD 20742-4111}

\begin{abstract}
We examine the shift in the chiral condensate due to a constant
electromagnetic field at ${\cal O}(p^6)$ using $SU(2)$ chiral
perturbation theory and a realistic $M_\pi = 140 \textrm{ MeV}$. We
find that this value differs significantly from the value calculated
using $M_\pi = 0$, while the magnitude of the two-loop correction is
unclear due to the uncertainty in the experimentally determined value
of the relevant ${\cal L}_6$ LEC.
\end{abstract}

\maketitle

\section{Introduction}

In QCD, the chiral condensate is important as it is the order
parameter of chiral symmetry breaking. As such, its behavior is
key in understanding QCD in extreme conditions. Finite temperature
and pressure effects have been studied extensively. On the other
hand, comparatively few efforts have focused on the effects of
finite electromagnetic fields. Prior studies have been done either
using chiral perturbation theory
($\chi$PT)\cite{Smilga,Shushpanov,GasserLeutwyler}, the effective
theory for low-energy QCD, or models compatible with the
large-$N_C$ expansion\cite{Gorbar,Goyal} (the NJL model, in
particular\cite{Klevansky,NJL,SmilgaFurther,Gusynin}). The method
used in this paper is $\chi$PT. This approach has the advantage of
being model-independent and systematic, but has the disadvantage
of containing a number of undetermined parameters (low-energy
constants or LECs).  We extend prior work done at the $M_\pi=0$
limit \cite{Smilga,Shushpanov} and at one loop \cite{CMW} to two
loops for $M_\pi = 140\textrm{ MeV}$.

Other calculations using $\chi$PT have focused on $M_\pi = 0$, and
while this may be of interest theoretically, it has at best a
narrow window of validity\cite{CMW}. In addition, it is unlikely
to find in nature a real electric or magnetic field with $eE \gg
M_\pi^2$ or $eH \gg M_\pi^2$, which is required for the
approximation to be reasonable. Incidentally, the opposite limit,
$M_\pi^2 \gg eH$, might be of more interest in the sense that it
can be produced in the laboratory, but in such a regime the shift
in the condensate is miniscule.

Continuing a calculation to higher orders in an expansion is
always of interest at least in the trivial sense of finding a more
precise result. In this case, large-$N_C$ QCD\cite{tHooft,Witten}
provides another possible motivation for why the ${\cal O}(p^6)$
result might be of interest. Low-energy constants (LECs) of the
same chiral order will have different orders of $N_C$, depending
upon the number of flavor traces in the term they multiply. This
can be understood as follows: a flavor trace corresponds to a
quark loop in the analogous QCD calculation, and large-$N_C$
counting rules indicate quark loops are down by a power of $N_C$,
and the LEC is the only parameter available to absorb this
difference.

This large-$N_C$ dependence of the LECs can provide hints to the
convergence of the chiral expansion. There are several processes which
have been calculated at two-loop order, most of which show a close
match between the ${\cal O}(p^6)$ results and the experimental results
\cite{BijnensBeyond}. The ${\cal O}(p^6)$ correction required to
achieve this agreement varies, but in particular, the ${\cal O}(p^6)$
calculation of the process $\gamma\gamma \to \pi^0\pm^0$ is strikingly
more accurate than the ${\cal O}(p^4)$ result\cite{Bellucci}. This is
also the process which provides a (very rough) estimate of the LEC we
use here, so it is conceivable that the ${\cal O}(p^6)$ correction
will be important in our case, as well.

We will first proceed with a brief overview of $SU(2)$ $\chi$PT. We
then follow with an analytical calculation of the shift in the chiral
condensate due to a magnetic field at ${\cal O}(p^6)$, and finally, a
numerical analysis of the shift for general electromagnetic fields.

\section{Chiral Perturbation Theory}

\subsection{Basics}

This section is a brief summary of the chiral perturbation theory
notation that will be used in this paper. For a detailed
description of the theory, see the original papers by Gasser and
Leutwyler\cite{GasserLeutwyler} or a number of
reviews(refs.~\cite{Pich,Ecker}, for example). Treatment of the
chiral lagrangian to ${\cal O}(p^6)$ can be found in
\cite{Bijnens,BijnensR,BijnensBeyond}.

The building blocks that are used to construct the chiral lagrangian
include $U = u(\phi)^2$, containing the dynamical pion fields, and
external fields $s$, $p$, $a_{\mu}$, and $v_{\mu}$. We will consider here
$SU(2)$ flavor symmetry with $m_u \approx m_d$, with external fields
corresponding to only a constant electromagnetic field and quark
masses. We then have
\begin{equation}
\begin{split}
\chi = 2B(s + i p) = 2B{\cal M}
\\r_\mu = l_\mu = -eQA_\mu = -e \left(\frac{\tau_3}{2}\right) A_\mu.
\end{split}
\end{equation}
We also need to define the covariant derivative as
\begin{equation}
D_\mu U = \partial_\mu U - i r_\mu U + i U l_\mu.
\end{equation}
From these, we define the following operators that will contribute
to the terms relevant in this paper (in the general $SU(n)$ notation):
\begin{equation}
\begin{split}
u_\mu = & i \{u^\dag (\partial_\mu - i r_\mu)u-u(\partial_\mu - i l_\mu)u^\dag\}
\\\chi_\pm = &u^\dag\chi u^\dag \pm u\chi^\dag u
\\f_\pm^{\mu\nu} = &u F_L^{\mu\nu} u^\dag \pm u^\dag F_R^{\mu\nu} u
\\\chi_-^\mu = &u^\dag D^\mu\chi u^\dag - u D^\mu \chi^\dag u
\end{split}
\end{equation}
where are defined
\begin{equation}
\begin{split}
F_R^{\mu\nu} = \partial^\mu r^\nu - \partial^\nu r^\mu - i[r^\mu,r^\nu]
\\F_L^{\mu\nu} = \partial^\mu l^\nu - \partial^\nu l^\mu - i[l^\mu,l^\nu].
\end{split}
\end{equation}
Note that in the case we are considering, $F_R^{\mu\nu} =
F_L^{\mu\nu}$.

$U$ can be paramaterized in several ways, but we will use the Weinberg parameterization
\begin{equation}
U=\sigma + \frac{i \pi^a\tau^a}{F}, \sigma^2 + \frac{\vec{\pi}^2}{F^2} = 1.
\end{equation}
Here, $\pi^a$ are still the dynamical fields, and $\sigma$ is
represented as an expansion in terms of $\pi^a$ from the second
equation.

Using these definitions, ${\cal L}_2$ and ${\cal L}_4$ can be written
as follows (in the general $SU(N)$ form), where $\langle A\rangle$ denotes the
trace of $A$ \cite{Bellucci}:
\begin{equation}
\begin{split}
{\cal L}_2 = & \frac{F^2}{2}\langle u_\mu u^\mu + \chi_+\rangle
\\{\cal L}_4 = & \frac{l_1}{4}\langle u^\mu u_\mu\rangle^2 + \frac{l_2}{4}\langle u_\mu u_\nu \rangle \langle u^\mu u^\nu \rangle + \frac{l_3}{16}\langle \chi_+ \rangle^2
\\&+ \frac{i l_4}{4}\langle u_\mu \chi_-^\mu\rangle - \frac{l_5}{2}\langle f_-^{\mu\nu}f_{-\mu\nu}\rangle
\\&+ \frac{i l_6}{4}\langle f_+^{\mu\nu}\left[u_\mu,u_\nu\right]\rangle - \frac{l_7}{16}\langle \chi_- \rangle^2
\\&+ \textrm{contact terms}
\end{split}
\end{equation}

The calculation here is up to ${\cal O}(p^6)$, which means that we
will be using the ${\cal L}_2$ lagrangian up to two loops and the
${\cal L}_4$ lagrangian up to one loop; the ${\cal L}_6$ lagrangian
will also contribute at tree level. This lagrangian has been
calculated in ref.~\cite{Bijnens}, and has more terms than we will
list (112 for $SU(n)$ and 53 for $SU(2)$). Fortunately, only one of
these (in $SU(2)$) will be relevant for our calculation, as we will
see later, and it can be expressed as:
\begin{equation}
{\cal L}_6 = c_{34} \langle \chi_+ f_{+ \mu\nu}f_+^{\mu\nu}\rangle  + \sum_{i\neq34}c_i P_i
\end{equation}

\subsection{Renormalization}

Renormalization of the theory to ${\cal O}(p^4)$ was calculated in
ref.~\cite{GasserLeutwyler}. It has also more recently been calculated
for the ${\cal O}(p^6)$ lagrangian in \cite{BijnensR}.

Using the $SU(2)$ LECs, the renormalized couplings can be written:
\begin{equation}
\begin{split}
l_i & = (c\mu)^{d-4}\left(l_i^r + \gamma_i \Lambda\right)
\\\Lambda & = \frac{1}{16\pi^2(d-4)}
\\\gamma_1 & = \frac{1}{3}, \gamma_2 = \frac{2}{3}, \gamma_3=-\frac{1}{2}, \gamma_4 = 2,
\\\gamma_5 & = -\frac{1}{6}, \gamma_6 = -\frac{1}{3}, \gamma_7 = 0
\end{split}
\end{equation}
When $M \neq 0$, these can be expressed in terms of scale-independent parameters as:
\begin{equation}
l_i^r = \frac{\gamma_i}{32\pi^2}\left(\bar{l}_i+\log\frac{M^2}{\mu^2}\right)
\end{equation}
The renormalization of the ${\cal L}_6$ term that we will be using
later can be expressed similarly in terms of the renormalized LECs from ${\cal L}_4$ as
\begin{equation}
\begin{split}
c_i & = \frac{(c\mu)^{2(d-4)}}{F^2}\left(c_i^r(\mu,d) - \gamma_i^{(2)}\Lambda^2
   - (\gamma_i^{(1)}+\gamma_i^{(L)}(\mu,d))\Lambda\right)
\\\gamma_{34}^{(L)} & = -l_5^r + \frac{1}{2}l_6^r, \gamma_{34}^{(1)} = \gamma_{34}^{(2)} = 0
\end{split}
\end{equation}

\section{Calculation of $\Sigma$ from vacuum energy}

The term in the QCD lagrangian which is relevant to $\Sigma$ is $m_q
\bar{q}q$.  Since we know that $\Sigma\sim \langle \bar{q}{q}\rangle$,
in the isospin limit of $m_u = m_d = \hat{m}$ we can calculate the
condensate from the vacuum energy as
\begin{equation}
\Sigma = - \frac{\partial \epsilon_\textrm{vac}}{\partial\hat{m}}.
\end{equation}
To first order in $M_\pi^2/F_\pi^2$, the Gell-Mann-Oakes-Renner
relation $F_\pi^2M_\pi^2 = \Sigma(m_u+m_d)$\cite{GMOR} applies, and
can be used to calculate the shift in the condensate.  Unfortunately,
we will need the next order result, which will introduce some
ambiguity as follows. Neither the chiral condensate or the quark mass
can be defined independently; only their product has a physical
meaning. In the language of $\chi$PT\cite{GasserLeutwyler},
\begin{equation}
2\hat{m}\Sigma = F^2M^2\left\{1 + \frac{M_\pi^2}{32 \pi^2 F_\pi^2}(4\bar{h}_1-\bar{l}_3) + {\cal O}(M_\pi^4)\right\}.
\end{equation}
The ambiguity here is codified in the unphysical LEC $\bar{h}_1$,
which will vary according to the renormalization convention. This is
an ambiguity in the definition of $\Sigma$. In order to avoid this
difficulty, we choose to normalize our results according to the
quantity $\Sigma_0$, which we define by
\begin{equation}
2\hat{m}\Sigma_0 = F_\pi^2 M_\pi^2,
\end{equation}
with $F_\pi$ and $M_\pi$ at their physical values. We will thus
express our results in terms of $\Delta\Sigma/\Sigma_0$, where
$\Delta\Sigma \equiv \Sigma(H) - \Sigma(H=0)$. As we will see, this
ratio is unambiguous at the order to which we work.

We will also need the relationships between $M$ and $M_\pi$ as
well as $F$ and $F_\pi$ ($M^2 \equiv 2B\hat{m}$).  The difference
between the lowest-order ${\cal O}(p^4)$ result and the ${\cal
O}(p^6)$ result is ${\cal O}(p^2)$; we therefore only need one
order of corrections\cite{GasserLeutwyler}:
\begin{equation}
\begin{split}
\label{eqn:constcorrect}
M_\pi^2 = &M^2\left[1 - \frac{M^2}{32\pi^2F^2}\bar{l}_3 + {\cal O}(M^4)\right]
\\F_\pi = &F\left[1 + \frac{M^2}{16\pi^2F^2}\bar{l}_4 + {\cal O}(M^4)\right].
\end{split}
\end{equation}
This will only be applicable in the first term of the expansion, which
is two powers of momentum less than the maximum order for the
calculation.

Our object is to calculate the condensate for the case of a
constant electromagnetic field at ${\cal O}(p^6)$ in $\chi$PT. As
described above, we can accomplish this by calculating the vacuum
energy to the same order. As noted by ref.~\cite{Shushpanov}, this
calculation will involve two-loop diagrams with ${\cal L}_2$
vertices, one-loop diagrams with an ${\cal L}_4$ vertex, and
tree-level diagrams with an ${\cal L}_6$ vertex. Diagrams which
contribute to the vacuum energy will contain only external photon
lines coming from the constant EM field (which is the ``vacuum''
in this case). Insertions of the electromagnetic field at the
${\cal L}_2$ level are calculated as part of the propagator of the
$\pi^{\pm}$. Thus, the ${\cal O}(p^4)$ calculation corresponds
roughly to a closed single propagator and was calculated in
ref.~\cite{CMW}. Contributing diagrams must be dependent on the
electromagnetic field, either through a direct insertion or
through the propagator of the $\pi^{\pm}$.

\begin{figure}[tbp]
\subfigure[\label{subfig:L2p0ppm}]{\includegraphics{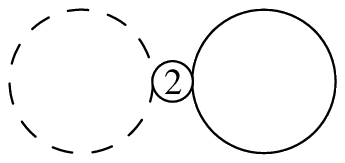}}
\hspace{0.02\textwidth}\subfigure[\label{subfig:L2ppmppm}]{\includegraphics{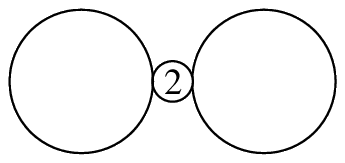}}
\\\subfigure[\label{subfig:L4ppmHH}]{\includegraphics{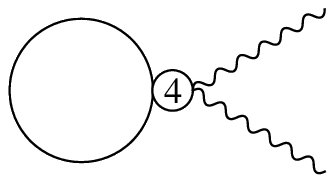}}
\hspace{0.02\textwidth}\subfigure[\label{subfig:L4ppm}]{\includegraphics{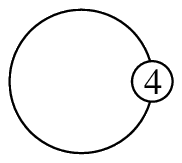}}
\hspace{0.02\textwidth}\subfigure[\label{subfig:L6HH}]{\includegraphics{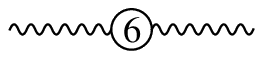}}
\caption{\label{fig:vacdiagrams}Diagrams contributing to the vacuum energy shift due to an electromagnetic field. Dashed lines denote $\pi^0$ and solid lines denote $\pi^\pm$.}
\end{figure}

With these criteria, we find that the diagrams contributing to our
calculation are as pictured in Fig.~\ref{fig:vacdiagrams}.  Returning,
then, to the $\chi$PT lagrangian, we find that only the terms
proportional to $l_3$, $l_4$, $l_5$, and $l_6$ can contribute from the
${\cal L}_4$ lagrangian, and only the term proportional to $c_{34}$
(as we anticipated above) can contribute to the vacuum energy from
${\cal L}_6$.

We simplify the chiral lagrangians for $SU(2)$ and $m_u = m_d$ up to
the relevant terms in ${\cal L}_6$\cite{Shushpanov}, including only
terms which will contribute to the diagrams in
Figure~\ref{fig:vacdiagrams}:
\begin{equation}
\begin{split}
{\cal L}_2 = & \frac{1}{2}(\partial_\mu\pi^0)^2 - \frac{M^2(\pi^0)^2}{2} - M^2\pi^+\pi^-
\\&+(\partial_\mu\pi^+ +ieA_\mu\pi^+)(\partial^\mu\pi^- -ieA^\mu\pi^-)
\\&+\frac{1}{2 F^2}\left[\pi^0\partial_\mu \pi^0 + \partial_\mu(\pi^+\pi^-)\right]^2
\\&- \frac{M\pi^2}{8 F^2}\left[2\pi^+\pi^- + (\pi^0)^2\right]^2
\\{\cal L}_4 = & -\frac{2 l_5}{F^2}(e F_{\mu\nu})^2\pi^+\pi^-
\\&-\frac{2il_6}{F^2}e F_{\mu\nu}\left[\partial^\mu\pi^-\partial^\nu\pi^++ieA^\mu\partial^\nu(\pi^+\pi^-)\right]
\\&-2l_3\frac{M^4}{F^2}\pi^+\pi^-
\\{\cal L}_6 = & 4 c_{34} M^2 (eF_{\mu\nu})^2
\end{split}
\end{equation}
Here, terms proportional to $l_3$ and $l_4$ have been added to the
lagrangian from ref.~\cite{Shushpanov}, which are down by an order of
$M^2$ but have the same overall chiral order.

As a first case, we will work with the case of pure magnetic fields,
where $(eF_{\mu\nu})^2 = 2(eH)^2$. This simplifies the calculations
and allows us to obtain an analytic result.  We will later generalize
to an arbitrary combination of $E$ and $H$ fields for numerical
analysis.

The propagator for a scalar particle in a constant $H$ field was
first calculated in ref.~\cite{Schwinger}, and here we use the
convenient form also used in ref.~\cite{Shushpanov}:
\begin{equation}
\begin{split}
D^{H}(x,y) = & \Phi(x,y)\int\frac{d^4k}{(2\pi)^4}e^{ik(x-y)}D^{H}(k)
\\D^{H}(k) =  & \int_0^\infty\frac{ds}{\cosh(eHs)}e^{-s\left(k_{\|}^2+k_{\bot}^2\frac{\tanh eHs}{eHs} + M^2\right)},
\end{split}
\end{equation}
where $\Phi(x,y) = \exp\{ie\int_y^x A_\mu(z)dz_\mu\}$, $k_{\|}^2 =
k_3^2+k_4^2$ and $k_{\bot}^2 = k_1^2+k_2^2$.

We will also need the scalar propagator
\begin{equation}
\begin{split}
D(0)\equiv & D(x,x) = \int\frac{d^dk}{k^2+M^2}
\\=& 2M^2(c\mu)^{d-4}\left[\Lambda + \frac{1}{32\pi^2}\log\frac{M^2}{\mu^2}\right]
\\\Lambda = & \frac{1}{16\pi^2(d-4)}.
\end{split}
\end{equation}
$D(0)$ and $D^H(0)\equiv D^H(x,x)$ are both divergent quantities,
whereas $D^H(0) - D(0)$ is finite:
\begin{equation}
\begin{split}
D^{\Delta H}(0) \equiv & D^H(0) - D(0)
\\= & -\frac{eH}{16\pi^2}\int_0^\infty\frac{dx}{x^2}e^{-\beta x}\left(1 - \frac{x}{\sinh x}\right)
\end{split}
\end{equation}
with $\beta = M^2/eH$. This is the same integral as was calculated in ref.~\cite{CMW}, and can be expressed analytically as:
\begin{equation}
\begin{split}
D^{\Delta H}(0) = & -\frac{eH}{16\pi^2} I_H(\beta)
\\I_H(\beta) = & \log(2 \pi) +\beta\log \left(\frac{\beta}{2}\right) - \beta
- 2 \log \Gamma\left(\frac{1 + \beta}{2}\right)
\end{split}
\end{equation}

With these, the diagrams in Fig.~\ref{fig:vacdiagrams} can be
calculated fairly straightforwardly to be \cite{Shushpanov}
\begin{equation}
\begin{split}
\epsilon_{\ref{subfig:L2p0ppm}}^{(2)} = & \frac{M^2}{2F^2}D(0)D^H(0)
\\\epsilon_{\ref{subfig:L2ppmppm}}^{(2)} = & \frac{1}{F^2}D^H(0)\int\frac{d^dk}{(2\pi)^d}(k^2 + M^2)D^H(k)
\\\epsilon_{\ref{subfig:L4ppmHH}}^{(2)} = & \frac{2 (eH)^2}{F^2}(2l_5-l_6)D^H(0)
\\\epsilon_{\ref{subfig:L4ppm}}^{(2)} = & 2l_3\frac{M^4}{F^2}D^H(0)
\\\epsilon_{\ref{subfig:L6HH}}^{(2)} = & -8 c_{34}M^2 (eH)^2
\end{split}
\end{equation}
The only one of these diagrams which is not expressed solely in terms
of $D^H(0)$ and $D(0)$ is $\epsilon_{\ref{subfig:L2ppmppm}}^{(2)}$,
which vanishes generally as well as in the $M_\pi = 0$ case.

In order to make the divergences and scale-dependence explicit, we
first make the substitution $D^H(0) = D^{\Delta H}(0) + D(0)$. Any
term which is dependent on neither $H$ nor $D^H(0)$ can then be
re-absorbed into the vacuum energy; we are only looking for the shift
due to $H$. After making this substitution, we see that
$\epsilon_{\ref{subfig:L2p0ppm}}^{(2)}$ is divergent and cancelled by
a counterterm generated by the combination $(2 l_3 +
l_4)$. $\epsilon_{\ref{subfig:L4ppmHH}}^{(2)}$ has both a finite
piece, which will contribute to the calculation, and a divergent
piece, which is cancelled by a counterterm in
$c_{34}$. $\epsilon_{\ref{subfig:L4ppm}}^{(2)}$ and
$\epsilon_{\ref{subfig:L6HH}}^{(2)}$ are finite, aside from the
aforementioned counterterms.

Combining, then, all of these terms, we find the vacuum energy to be
\begin{equation}
\begin{split}
\epsilon^{(2)}(H) = - \frac{(eH)^3}{(16\pi^2)^2 F^2}
\Bigg\{ & I_H(\beta)\left[\frac{1}{3}(\bar{l}_6-\bar{l}_5) - \frac{\beta^2}{2}\bar{l}_3\right]
\\&+ \beta \bar{d}(M^2)\Bigg\},
\end{split}
\end{equation}
where we have defined the scale-independent quantity
\begin{equation}
\bar{d}(M^2) = 8(16\pi^2)^2c_{34}^r - \frac{1}{3}(\bar{l}_6-\bar{l}_5)log\left(\frac{M^2}{\mu^2}\right).
\end{equation}
We then substitute Eq.~(\ref{eqn:constcorrect}) to find $M_\pi^2$
from $M^2$ in the first -order term. We find that the $\bar{l}_3$
term cancels, and that the correction to $F$ does not play a role
(as it only appears at the second order). Then, taking a
derivative, and applying the Gell-Mann-Oakes-Renner relation as
above, with the first-order corrections for the ${\cal O}(p^4)$
term, we find ($\beta_\pi = M_\pi^2/eH$):
\begin{widetext}
\begin{equation}
\label{eqn:Hshift}
\begin{split}
\frac{\Delta\Sigma(H)}{\Sigma_0}& = \frac{eH}{16\pi^2 F_\pi^2}I_H(\beta_\pi)
  +\left(\frac{eH}{16\pi^2 F_\pi^2}\right)^2
     \Bigg\{
        -\frac{1}{3}(\bar{l}_6-\bar{l}_5)\left[1+\log 2 +\psi\left(\frac{1+\beta_\pi}{2}\right)\right]
       +\bar{d}(eH)
     \Bigg\},
\end{split}
\end{equation}
\end{widetext}
with $\psi(x) \equiv \frac{d}{dx}\log\Gamma(x)$

Taking $\beta \to 0$, $\psi\left(\frac{1}{2}\right) = - \gamma_e$, and
the shift we find agrees with the expression found in
ref.~\cite{Shushpanov} for the case $M_\pi = 0$.

For the case of an $E$ field, we can make the substitution $H \to i E$
and get a similar analytic expression. The $E\cdot H \neq 0$ case is
somewhat more complicated, but we will write an integral expression
which we can later evaluate numerically.

For a convenient paramaterization of the general case, we introduce
the variables $\phi$ and $f$ such that with ${\cal F}=
\frac{H^2-E^2}{2} = \frac{1}{4}F_{\mu\nu}^2$ and ${\cal G }=\vec{E}
\cdot \vec{H}$ \cite{CMW},
\begin{equation}
{\cal F}= \frac{f^2 \cos (2 \phi) }{2}  \; \; \; {\cal G}= \frac{f^2
\sin (2 \phi) }{2}.
\end{equation}

Expressed in terms of these variables, the shift in the condensate due
to an arbitrary combination of fields will become ($\beta_f = M_\pi^2/ef$):
\begin{widetext}
\begin{equation}
\label{eqn:EHshift}
\begin{split}
\frac{\Delta\Sigma({\cal F},{\cal G})}{\Sigma_0} &=
\frac{ef}{16 \pi^2 F_\pi^2}I_{EH}(\beta_f,\phi) + \left(\frac{ef}{16 \pi^2 F_\pi^2}\right)^2\cos 2\phi\left\{\frac{1}{3}(\bar{l}_6-\bar{l}_5)(I_{EH}'(\beta_f,\phi) - 1) + \bar{d}(M_\pi^2)\right\}
\\I_{EH}(\beta_f,\phi) &= \int_0^\infty \frac{dz}{z^2}e^{-\beta_f z} \left[1 - \frac{z^2 \sin 2\phi}{2 \sin(z \sin\phi)\sinh(z\cos\phi)+i\epsilon}\right]
\end{split}
\end{equation}
\end{widetext}

This is the same integral as in ref.~\cite{CMW}, and as before we have
had to avoid some potential ambiguity. The poles in the integrand
indicate an instability in the system, which is interpreted as due to
pair creation in an electric field\cite{Schwinger}. We have chosen to
regulate the divergence in a manner which has an imaginary part
corresponding to this pair creation. The magnitude of the imaginary
part indicates the importance of this instability, though some caution
is warranted in interpreting it quantitatively. This issue was
discussed in more detail in ref.~\cite{CMW}.

Equations~(\ref{eqn:Hshift}) and (\ref{eqn:EHshift}) are the
principal results of this work.

\section{Numerical results}

Because the behavior of the theory is encoded in the LECs, a
real-world interpretation of the low-energy behavior requires the use
of measured LECs. For the ${\cal L}_4$ LECs, this is straightforward,
as these are individually determined with relatively small error. On
the other hand, ${\cal L}_6$ LECs, such as $c_{34}$, are more
problematic, as there are in general many more LECs than easily
measurable processes to determine them. These LECs are often estimated
(at a particular scale) by resonance exchange.  Unfortunately,
$c_{34}$ in particular is difficult to extract, as the resonance
processes to which it contributes involve only scalar exchange, and
because it appears squared in these processes, its sign is
undetermined. This resonance exchange occurs at a scale $~M_\rho =
768$\cite{Bellucci}, and it is the (scale-dependent) value determined
by experiment that has an undetermined sign. The scale-independent
$\bar{d}$ is positive in both cases.

The values we use for these constants are \cite{Bellucci,BijnensNumsOld}:
\begin{equation}
\begin{split}
\bar{l}_6-\bar{l}_5 &= 3.0 \pm 0.3
\\d^r \equiv 8(16\pi^2)^2c_{34}^r &= \pm 1.5\pm 1.5.
\end{split}
\end{equation}

\begin{figure}[tbp]
\includegraphics[width=0.99\linewidth,height=175pt]{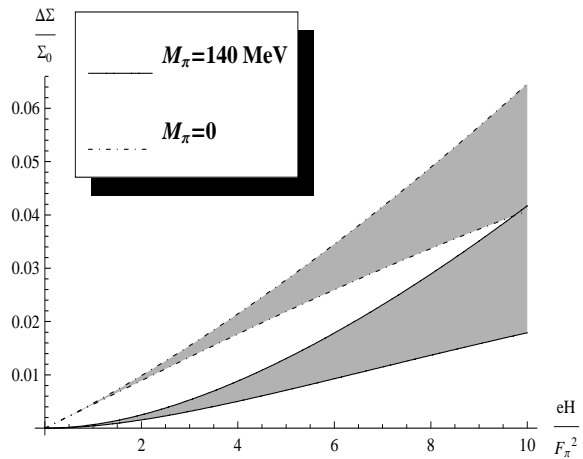}
\caption{\label{fig:shush}A comparison of the shift due to a pure
magnetic field in the $M_\pi=0$ case to the $M_\pi=140 \textrm{ MeV}$
case. Shaded regions indicate uncertainty due to the ${\cal L}_6$
constant $d^r$.}
\end{figure}

With these experimental values, we can plot realistic values of
the shift in the condensate. In Fig.~\ref{fig:shush}, for the case
of a pure magnetic field, we compare the value calculated for a
finite $M_\pi$ to that for $M_\pi=0$. It is clear that these
values are significantly different, as in the ${\cal O}(p^4)$
case\cite{CMW}. In this and the figures following, we have chosen
to extend our results up to $ef = 290\textrm{ MeV}$ (the expansion
parameter is $\Lambda=4\pi F_\pi = 1.2 \textrm{ GeV}$).

Here, we use the same numerical trick as in ref.~\cite{CMW} to extract
the principal value of the integral numerically. We remove the
singularities due to the poles located at $z_i$ with residue $R_i$ by
subtracting the expression
\begin{equation}
i \sum_{n} R_n(z_n) \left(\frac{1}{z - z_n}-\frac{1}{z + z_n}\right).
\end{equation}
The principal value of the integral of this expression is zero, but it
has a singularity at $z_i$ which exactly cancels the singular behavior
of the integrand in $I_{EH}$.

\begin{figure}[tbp]
\subfigure[\label{subfig:ReTot}]{\includegraphics[width=0.99\linewidth,height=175pt]{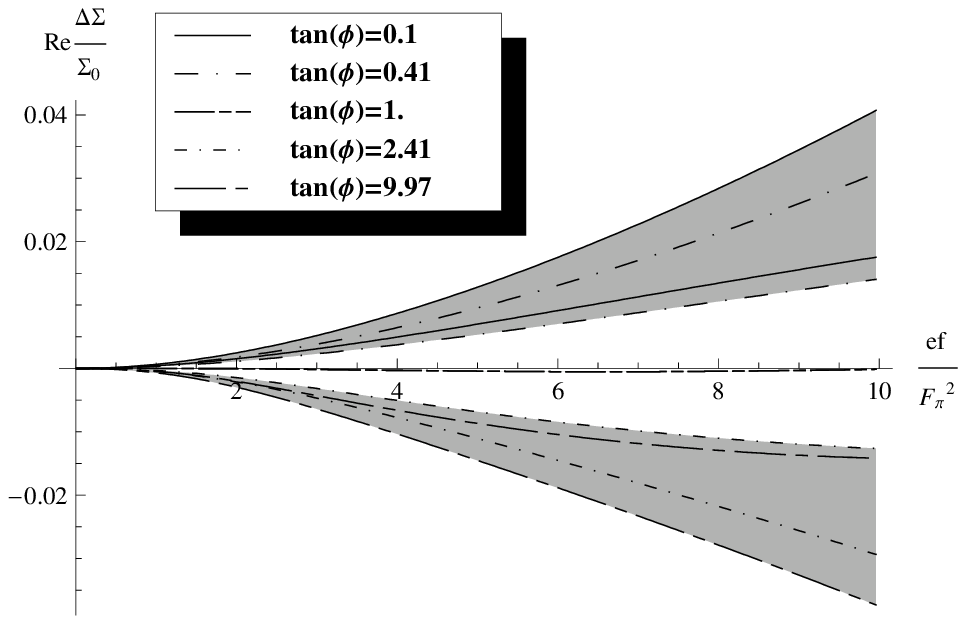}}
\subfigure[\label{subfig:ImTot}]{\includegraphics[width=0.99\linewidth,height=175pt]{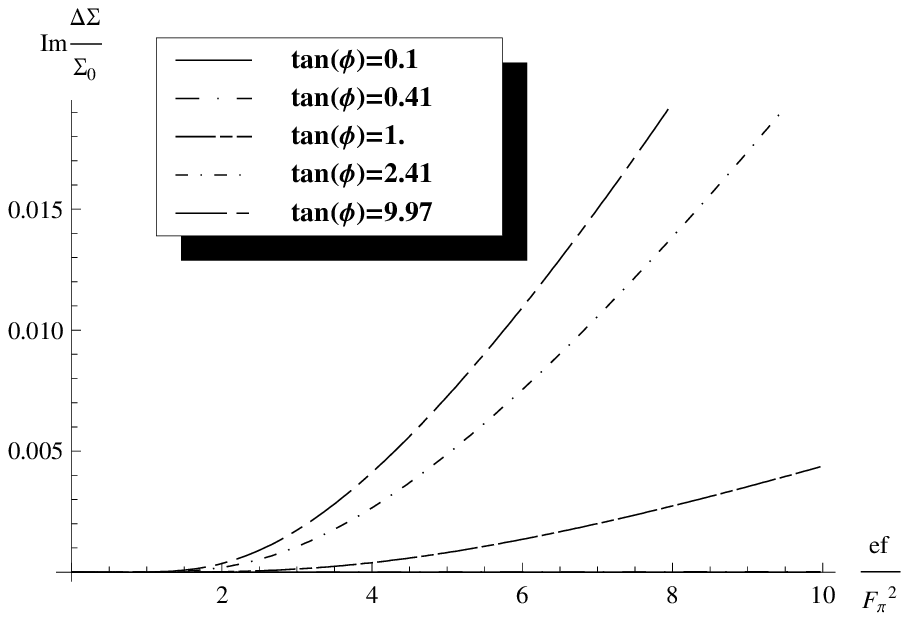}}
\caption{\label{fig:totval}The imaginary and real parts of the total value of the shift in the condensate due to general $E$ and $H$ fields, with $f$ and $\phi$ as defined in the text. Shading depicts uncertainty due to $c_{34}$.}.
\end{figure}

\begin{figure}[tbp]
\subfigure[\label{subfig:ReFrac}]{\includegraphics[width=0.99\linewidth,height=175pt]{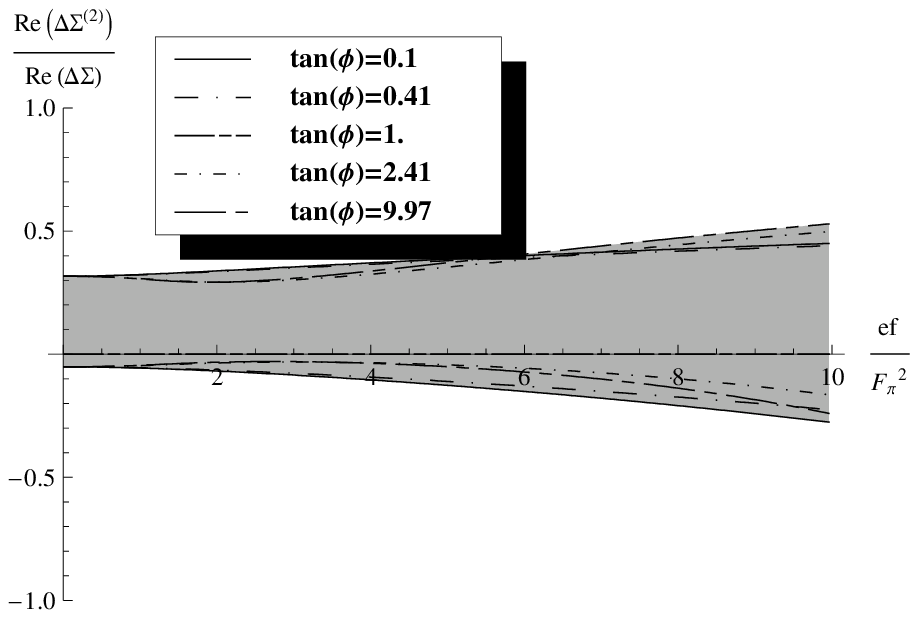}}
\subfigure[\label{subfig:ImFrac}]{\includegraphics[width=0.99\linewidth,height=175pt]{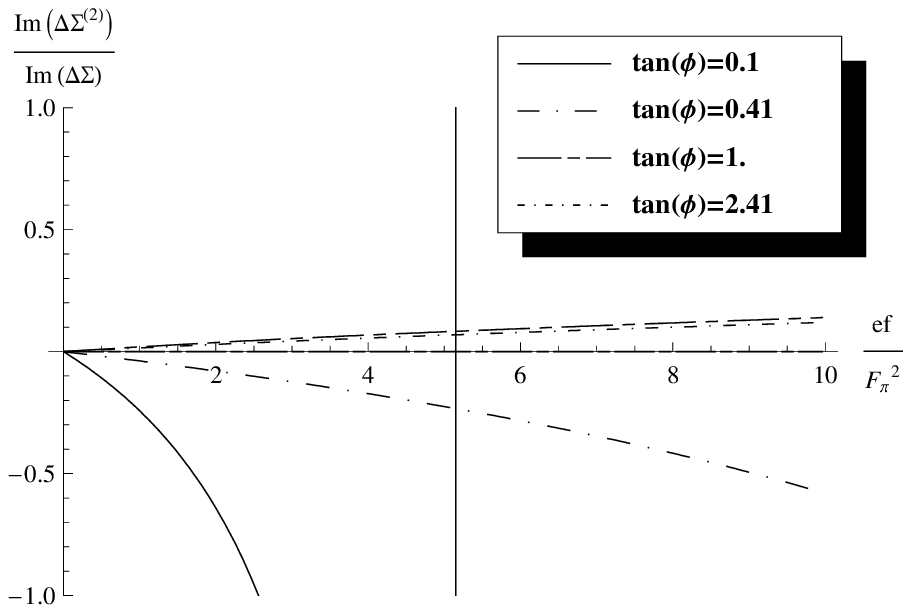}}
\caption{\label{fig:fracval}The imaginary and real parts of the ratio of the shift at two loops to the total shift for the case of general $E$ and $H$ fields, with $f$ and $\phi$ as defined in the text. Shading depicts uncertainty due to $c_{34}$.}.
\end{figure}

Using this method, we plot the total shift in the condensate up to
${\cal O}(p^6)$ from a general $E$ and $H$ field in
Fig.~\ref{fig:totval}, and in Fig.~\ref{fig:fracval}, we plot the
ratio of the added correction at ${\cal O}(p^6)$ to the total
shift. In these plots, we have included a shaded region to
indicate the possible values for the shift based on a range for
$d^r$ of $(-3,3)$.

The asymptotic expression for the shift as $\beta_\pi \to \infty$ (for
an $H$ field, which also provides some qualitative insight to other
cases) is
\begin{equation}
\label{eqn:asymshift}
\frac{\Delta\Sigma(H)}{\Sigma_0} = \frac{eH}{16\pi^2 F_\pi^2}\left(\frac{F_\pi^2}{6 M_\pi^2} - \frac{\bar{l}_6 - \bar{l}_5}{48\pi^2} + \frac{\bar{d}}{16\pi^2}\right),
\end{equation}
which is, of course, zero for $\beta_\pi\to\infty$ ($H\to 0$).
This expression encodes low-energy behavior for a more realistic
regime,; namely, that of the actual pion mass and a very small
magnetic field. We see from Fig.~\ref{fig:fracval} and in
(\ref{eqn:asymshift}) that the (unknown) sign of $d^r$ has a
profound impact on the importance of the ${\cal O}(p^6)$
calculation. The shift in the condensate for a positive $d^r$ is
significant enough even as $f\to 0$, whereas the shift for a
negative one is negligible up to large values of $ef/F_\pi^2$.

Another notable feature is that the contribution to the imaginary
part is larger at ${\cal O}(p^6)$ order as a pure $H$ field is
approached (while the total imaginary part is going to zero).
Except in this case where the total imaginary part is negligible,
the fraction of the imaginary part at two loops will be much less
significant than its real counterpart in regimes where the chiral
expansion would be expected to converge ($ef \sim M_\pi^2$ or
below).

The calculation as a whole will only be valid when the real part is
significantly larger than the imaginary part. When the imaginary part
dominates, the system will break down due to the instability from pair
creation.

\section{Discussion and Conclusions}

We have studied the shift in the chiral condensate due to an
electromagnetic field using chiral perturbation theory, which is a
powerful tool for analyzing the low-energy behavior of QCD. Our
analysis was done at ${\cal O}(p^6)$ with $M_\pi = 140 \textrm{
MeV}$. It is obvious that the inclusion of a nonzero pion mass greatly
affects the result.

The importance of the ${\cal O}(p^6)$ correction is less
clear. Large-$N_C$ reasoning coupled with the results of model-based
calculations give circumstantial evidence that it could play an
important role in the final result. However, because the sign of the
relevant LEC at ${\cal L}_6$ is undetermined by experiment, its effect
at ${\cal O}(p^6)$ could be significant or virtually irrelevant.

\section{Acknowledgements}

I would like to thank Tom Cohen for many useful discussions.  This
research is funded by the U. S. Department of Energy under grant
number DE-FG02-93ER-40762.

\clearpage

\end{document}